\documentclass[12pt]{article}
\usepackage{amsmath,amsfonts}
\usepackage[nosort]{cite}
\usepackage{graphicx}
\usepackage{feynmp}
\newcommand{\be}{\begin{equation}}
\newcommand{\ee}{\end{equation}}
\newcommand{\bea}{\begin{eqnarray}}
\newcommand{\eea}{\end{eqnarray}}

\newcommand{\quarter}{\frac{1}{4}}
\newcommand{\vev}[1]{{\left< {#1} \right>}}

\newcommand{\abs}[1]{{\left| {#1} \right|}}
\newcommand{\eqn}[1]{(\ref{#1})}
\newcommand{\nn}{\nonumber}

\DeclareMathOperator{\Tr}{Tr}

\newcommand{\cN}{{\mathcal N}}

\newcommand{\cR}{{\mathcal R}}
\newcommand{\cO}{{\mathcal O}}

\newcommand{\cX}{{\mathcal X}}
\newcommand{\cY}{{\mathcal Y}}
\newcommand{\cZ}{{\mathcal Z}}
\newcommand{\cRNf}{{\mathcal R}_{{\mathcal N}=4}}

\renewcommand{\title}[1]{\vbox{\center\LARGE{#1}}\vspace{5mm}}
\renewcommand{\author}[1]{\vbox{\center#1}\vspace{5mm}}
\newcommand{\address}[1]{\vbox{\center\em#1}}
\newcommand{\email}[1]{\vbox{\center\tt#1}\vspace{5mm}}

\begin{document}
\begin{titlepage}
\begin{center}
\vspace{5mm}
\hfill {\tt HU-EP-08/62}\\
\vspace{20mm}
\title{The structure of $n$-point functions of chiral primary operators in ${\cal N}=4$ super Yang-Mills
at one-loop}
\author{\large Nadav Drukker and Jan Plefka}
\address{Institut f\"ur Physik, Humboldt-Universit\"at zu Berlin,\\
Newtonstra{\ss}e 15, D-12489 Berlin, Germany}

\email{drukker, plefka@physik.hu-berlin.de}

\end{center}

\abstract{
\noindent
We develop a compact representation of the one-loop $n$-point functions of 
all chiral primary operators in planar $SU(N)$, ${\cal N}=4$ super Yang-Mills theory 
in terms of tree-level disk correlation functions and the
scalar one-loop box integral.
As a check, known results for all four-point functions and for $n$-point extremal 
and near-extremal correlators are rederived.
The result is then used to evaluate explicitly a selection of five and six-point functions.
Our findings suggest that a general one-loop five-point function may be represented through the  
minimal four-point and five-point functions of weight two operators.}

\vfill

\end{titlepage}


\section{Introduction and conclusions}

The string-gauge theory duality in its best understood and most symmetric
formulation identifies ${\cal N}=4$ supersymmetric Yang-Mills theory with type IIB superstrings
on an $AdS_5\times S^5$ background \cite{AdSCFT}. 
Great progress in our understanding of this $AdS$/CFT
system has been achieved in recent years by exploiting its powerful superconformal symmetry
together with the discovered integrable \cite{AdS/INT} structures. From the gauge theory 
perspective the set of local observables
is given by $n$-point functions of gauge invariant composite operators 
which fall into multiplets of the superconformal symmetry group. In the 
early days of the $AdS$/CFT correspondence the study
of so called chiral primary operators, which are scalar composite operators in the symmetric
traceless representation of the $SU(4)$ R-symmetry group with Dynkin labels $[0,k,0]$, received
great attention. These BPS operators are annihilated by one-half of the Poincare supercharges 
and are dual to the infinite tower of Kaluza-Klein modes of the
$AdS_5\times S^5$ compactification of IIB supergravity. The two and three-point functions
of chiral primary operators are protected against radiative corrections \cite{3pt}, hence
the scaling dimensions are given by $\Delta=k$ exactly. And indeed  these free field theory 
correlation functions can be matched to the dual supergravity result \cite{Lee:1998bxa}. 
Since 2002 the interest has advanced to the true string theoretical domain of the 
correspondence in which string
excitations of the $AdS_5\times S^5$-theory are matched to generic local gauge theory
operators with nontrivial two-point functions, i.e.~anomalous scaling dimensions. 
Here one is close to a complete solution of the problem of determining the spectrum of scaling
dimensions in the theory, determined by a set of Bethe equations \cite{Betheeq} 
(see \cite{intrevs} for reviews). In parallel and connected to this, our 
understanding of scattering amplitudes in the theory has increased
considerably. Here a novel dual superconfromal symmetry of the theory 
in momentum space enabled one to find the all-loop form of maximally 
helicity violating gluon four and five-point amplitudes \cite{GluonScattering}. 
Moreover, these amplitudes are dual to light-like Wilson loops 
(see \cite{AldayRoiban} for a reviews).

Despite the protectedness of two and three-point functions of ${\cal N}=4$ super 
Yang-Mills chiral primaries, the $n\geq 4$ point functions  are in general 
highly non-trivial functions of the 't Hooft coupling constant $\lambda$ and $1/N$. 
A lot of work has been devoted to the study of four-point functions, both from the 
gauge theory side up to two-loop order and in the supergravity approximation 
\cite{Eden:1998hh,D'Hoker:1999pj,D'Hoker:1999ni,Eden:1999kh,Arutyunov:2000py,
Dolan:2000ut,Heslop:2002hp, Arutyunov:2002fh,Arutyunov:2003ae,Arutyunov:2003ad,
Dolan:2004mu, Nirschl:2004pa}.%
\footnote{It should be noted that there exists a very specific class of so called 
extremal and next-to-extremal $n$-point functions, which are characterized by the 
property of having tree-level diagrams factorized into two and three-point 
functions. These $(n\geq 4)$-point functions do not 
receive one-loop corrections \cite{extremal} as well.} 
An important structural insight has been the universal 
factorization of the quantum  corrections to the correlation 
function of four chiral primaries of weight $k$
into a universal prefactor ${\cal R}_{{\cal N}=4}$ and a non-universal 
remainder \cite{Arutyunov:2002fh}
\be
\langle \cO_{k}(x_1)\, \cO_{k}(x_3)\, \cO_{k}(x_3)\, \cO_{k}(x_4)\, \rangle_{\text{quant}}
= {\cal R}_{{\cal N}=4}(s,t)\cdot {\cal F}^{(k)}(s,t,\lambda) \,,
\label{eins}
\ee
with conformal cross-ratios $s$ and $t$. This factorization was shown 
to arise both at weak and strong coupling, as well as non-perturbatively in 
instanton computations \cite{Arutyunov:2000im}. The prefactor 
$\cR_{\cN=4}$ was analyzed also in \cite{Dolan:2004mu, Nirschl:2004pa}, 
where it was related to 
superconformal Ward-Takahashi identities satisfied by the 4-point correlation 
functions.

In this note we wish to modestly extend these structural insights to $n\geq 4$ point functions of chiral 
primaries by performing a diagramatical computation at one-loop order in the
planar limit. 
The result of this paper is a universal form for the $n$-point function at one-loop
detailed in equation~\eqn{4-pt}: The one-loop contribution may be rewritten as a sum over 
certain tree-level amplitudes with the topology of the disc where $4$ insertions 
are at the boundary and $(n-4)$ insertions are situated in 
the bulk, multiplied by the one-loop scalar box integral.
The factorization with ${\cal R}_{{\cal N}=4}$ for $n=4$ of equation~\eqn{eins} 
at one-loop order is an immediate corrollary of our
result. In essence our result is a consequence of the fact that the one-loop interactions
involve at most four points and that the conformal symmetry does not allow for the appearance
of non-trivial functions depending on only three space-time points.

Furthermore, we apply this result to compute a selection of five and six-point functions. 
The structures we find suggest that the one-loop corrections to a general five-point 
function may be decomposed into the two minimal four-point and five-point functions of weight 
two operators multiplied by tree-level contractions.

The motivation for these considerations comes from at least two viewpoints: 
In a companion paper \cite{us2}, we study examples of $n$-point functions 
of chiral primary operators who have common supersymmetries. We apply 
the results of this paper to evaluate the one-loop corrections to these 
$n$-point functions and find that the radiative corrections vanish. Beyond this 
specific appication, we feel that following the tremendous advances on our 
understanding of $n=2$ point functions in the ${\cal N}=4$ gauge theory 
upon exploiting integrability, the time is ripe to turn one's attention to the 
cases with $n\geq 3$. We hope that our result will prove useful for such
an endeaviour in the future.

\section{Notation}

A chiral primary operator is a composite scalar field of dimension $k$  
being the lowest component of a $1/2$ BPS multiplet.
In terms of the elementary fields of ${\cal N}=4$ super Yang-Mills ($\phi^I,A_\mu,\psi^a$)
it is given by $\Tr[\,\phi^{\{I_1}\ldots \phi^{I_k\}}\, ]$ with $I=1,\ldots, 6$ and $\{\ldots \}$ denoting traceless symmetrization. A very convenient way to handle the $SO(6)$ indices is to 
represent the chiral primary with the help of a complex null vector $u^I$ (with $u^I\, u^I=0$ and
$u^I\,\bar u^I=1$) (see e.g.~\cite{Arutyunov:2002fh})
\be
\cO^{u}_{k}(x) := u^{I_1}\ldots  u^{I_k}\, \Tr\left[ \phi^{I_1}(x)\ldots \phi^{I_k}(x)\right] =
\Tr\left[ (u\cdot \phi(x))^k\right].
\ee
We shall be interested in computing $n$-point correlation functions of these operators
\be
\vev{\cO^{u_1}_{k_1}(x_1)\, \cO^{u_2}_{k_2}(x_2)\, \cO^{u_3}_{k_3}(x_3)\ldots
\cO^{u_n}_{k_n}(x_n)}
\label{objectofdesire}
\ee  
at one-loop. We furthermore define the tree level contraction of two scalars%
\footnote{Since we consider only planar diagrams, we suppress throughout 
the gauge group indices. Some of our considerations can be applied also 
to non-planar graphs.}
\be
[ij]:=\vev{(u_i\cdot \phi(x_i))\, (u_j\cdot \phi(x_j))}_{\text{tree}} = 
\frac{u_i\cdot u_j}{(2\pi)^2\, x_{ij}^2}\,, \qquad x_{ij}:= x_i-x_j \,.
\ee
Moreover, for any choice of four operators there are two conformally invariant 
cross-ratios, which will show up in the calculation. Focusing on the 
four-point function, we define $s$ and $t$ as the cross-ratios of 
$(x_1,x_2,x_3,x_4)$. They may also be expressed via one complex 
number $\mu$
\be
s=\frac{x_{12}^2\, x_{34}^2}{x_{13}^2\, x_{24}^2} =\mu\,\bar\mu\,  \qquad
t=\frac{x_{14}^2\, x_{23}^2}{x_{13}^2\, x_{24}^2} = (1-\mu)\, (1-\bar\mu)\,.
\label{cross-ratios}
\ee
With the definitons
\be
{\cal X}= [12][34]\,,\quad {\cal Y} =[13][24]\,, \quad {\cal Z}=[14][23]\,.
\label{XYZ}
\ee
the universal polynomial prefactor of equation~\eqn{eins} may be written
in two ways as
\begin{align}
{\cal R}_{{\cal N}=4} & =s\, (\cY-\cX)\, (\cZ-\cX) + t\, (\cZ-\cX)\, (\cZ-\cY) +
(\cY-\cX)\, (\cY-\cZ) \nonumber\\
&=\big(\mu\,(\cX-\cZ)+\cZ-\cY\big)\, 
\big(\bar\mu\,(\cX-\cZ)+\cZ-\cY\big)\,.
\label{cRN4}
\end{align}
Such factorization was also observed in \cite{Dolan:2004mu, Nirschl:2004pa}.

\section{The perturbative computation}

\begin{fmffile}{plane-graphs}

We present here some basic formulas that are helpful in order to 
address the one-loop radiative corrections to correlation functions of
local operators built from scalars.

Following \cite{Beisert:2002bb} we introduce the scalar propagator and some 
fundamental tree functions in configuration space
\begin{align}
I_{12} &=\frac{1}{(2\pi)^2(x_1-x_2)^2},  \label{I}\\
Y_{123}&=\int d^4 w \,I_{1w} I_{2w} I_{3w}, \label{Y}
\\
X_{1234}&=\int d^4 w \,I_{1w} I_{2w} I_{3w} I_{4w}, \label{X}
\\
H_{12,34}&=\int d^4 u \,d^4 v\, I_{1u} I_{2u} I_{uv} I_{3v} I_{4v}.
\label{H}
\end{align}
We have put the space-time points as indices to the function
to make the expressions more compact. These functions are
all finite except in certain limits. For example $Y$, $X$ and $H$
diverge logarithmically when $x_1\to x_2$.
The functions $X$ and $Y$ can be evaluated explicitly \cite{David}
\begin{align}
X_{1234}
&=\frac{\pi^2\Phi(s,t)}{(2\pi)^8 (x_1-x_3)^2(x_2-x_4)^2},
\\
Y_{123}&= \lim_{x_4\to\infty} (2\pi)^2\, x_4^2\, X_{1234}.
\end{align}
In the euclidean region
($\sqrt{s}+\sqrt{t}\geq 1$, $\abs{\sqrt{s}-\sqrt{t}}\leq 1$)
the function $\Phi(s,t)$ can be written in a manifestly real fashion as
\be
\begin{aligned}
\Phi(s,t)&=
\frac{1}{A}\,\text{Im}\left(\text{Li}_2 \frac{e^{i\varphi} \sqrt{s}}{\sqrt{t}}+
\ln\frac{\sqrt{s}}{\sqrt{t}}\,\ln\frac{\sqrt{t}-e^{i\varphi} \sqrt{s}}{\sqrt{t}}\right)
\\
e^{i\varphi}&= i\sqrt{-\frac{1-s-t-4i A}{1-s-t+4i A}},
\qquad
A=\quarter\sqrt{4st-(1-s-t)^2}.
\end{aligned}
\label{Phi}
\ee
It is positive everywhere, vanishes only in the limit
$s,t\to \infty$ and
has the hidden symmetry $\Phi(s,t)=\Phi (1/s,t/s)/s$.

There seems to be no analytic expression for the function $H$.
However, the Feynman rules lead to its appearance in the combination which 
can be expressed in terms of the functions $X$ and $Y$ 
\cite{Eden:1999kh,Beisert:2002bb}
\begin{align}
F_{12,34}&=
\frac{(\partial_1-\partial_2)\cdot (\partial_3-\partial_4) H_{12,34}}{I_{12}I_{34}}
=\frac{X_{1234}}{I_{13}I_{24}}
-\frac{X_{1234}}{I_{14}I_{23}}
+G_{1,34}-G_{2,34}+G_{3,12}-G_{4,12}\,,
\\
G_{1,34}&=\frac{Y_{134}}{I_{14}}-\frac{Y_{134}}{I_{13}}\,.
\end{align}

\subsection{Combining the basic interactions}

In order to simplify the one-loop perturbative computation of $n$-point functions of
chiral primaries of ${\cal N}=4$ super Yang-Mills we shall develop a 
number of insertion formulas, which allow one to easily construct the 
one-loop corrections to a given tree level graph.

We work  with scalar propagators normalized to
\be
[12]:=
u_1^{I_1}\, u_2^{I_2}\,\vev{\phi^{I_1}(x_1)\, \phi^{I_2}(x_2)}_{\text{tree level}} =
\raisebox{0mm}[7mm][7mm]{\parbox{4mm}{
\begin{fmfgraph*}(4,7) \fmfstraight
        \fmftop{a}
        \fmfbottom{b}
        \fmflabel{$u_1$}{a}
        \fmflabel{$u_2$}{b}
        \fmfdot{a,b}
        \fmf{plain,width=1}{a,b}
\end{fmfgraph*}}}
= (u_1\cdot u_2) \, I_{12}
 \label{normalization} \,.
\ee
As mentioned, we do not record $SU(N)$ factors.

Using this language we then note the following one-loop planar insertion formulas
\begin{align}
\quad
{\parbox{17mm}{
\begin{fmfgraph*}(10,10) \fmfstraight
        \fmfleft{a}
        \fmfright{b}
        \fmfdot{a,b}
        \fmflabel{$u_1$}{a}
        \fmflabel{$u_2$}{b}
        \fmf{plain,width=1}{a,v}
        \fmf{plain,width=1}{v,b}
        \fmfblob{.3w}{v}
\end{fmfgraph*}}}
& =\, -\lambda\,  (u_1\cdot u_2)\, I_{12}\, \frac{Y_{112}+Y_{122}}{I_{12}}
\label{seinsertion}\\
\raisebox{0mm}[10mm][10mm]{\parbox{17mm}{
\begin{fmfgraph*}(10,10) \fmfstraight
        \fmfright{u4,u2}
        \fmfleft{u3,u1}
        \fmfdot{u1,u2,u3,u4}
        \fmflabel{$u_2$}{u2}
        \fmflabel{$u_4$}{u4}
        \fmflabel{$u_1$}{u1}
        \fmflabel{$u_3$}{u3}
        \fmf{plain,width=1}{u1,v1,u2}
        \fmf{plain,width=1}{u3,v2,u4}\fmffreeze
        \fmf{wiggly}{v1,v2}
\end{fmfgraph*}}}
& = \,
 \frac{\lambda}{2}\, (u_1\cdot u_2)\, (u_3\cdot u_4)\, I_{12}\, I_{34}\, F_{12,34}
\label{Ginsertion} \\
\raisebox{0mm}[10mm][6mm]{\parbox{17mm}{
\begin{fmfgraph*}(10,10) \fmfstraight
        \fmftop{u1,u2}
        \fmfbottom{u3,u4}
        \fmflabel{$u_2$}{u2}
        \fmflabel{$u_4$}{u4}
        \fmflabel{$u_1$}{u1}
        \fmflabel{$u_3$}{u3}
        \fmfdot{u1,u2,u3,u4}
        \fmf{plain,width=1}{u1,v,u4}
        \fmf{plain,width=1}{u3,v,u2}
\end{fmfgraph*}}}
& =\,
\frac{\lambda}{2}\,\Bigl [\, 2\,(u_2\cdot u_3)\,(u_1\cdot u_4) -(u_2\cdot u_4)\,(u_1\cdot u_3)
-(u_1\cdot u_2)\,(u_3\cdot u_4) \Bigr ]\, X_{1234}
\label{Xinsertion}
 \end{align}
where the grey blob stands for the one-loop self-energy and curly lines 
denote gluon propagators. We furthermore note the relevant pinching 
limits of the functions defined in equations~\eqn{I}-\eqn{H} in point-splitting 
regularization
\be
\begin{aligned}
Y_{112} &= Y_{122}= -\frac{1}{16\pi^2}\left( \ln \frac{\epsilon^2}{x_{12}^2}-2\right)\, I_{12} \\
X_{1123} &= -\frac{1}{16\pi^2}\, I_{12}\, I_{13}\left( \ln \frac{\epsilon^2\, x_{23}^2}{x_{12}^2\,
x_{31}^2} -2 \right) \\
F_{12,13} &= -\frac{1}{16\pi^2}\left( \ln\frac{\epsilon^2}{x_{23}^2}  - 2\right)
+ Y_{123}\left(\frac{1}{
I_{12}} +\frac{1}{I_{13}} -\frac{2}{I_{23}}\right) \\
X_{1122} &= -\frac{1}{8\pi^2}\, I_{12}^2\left( \ln \frac{\epsilon^2}{x_{12}^2} -1 \right) \\
F_{12,12} &= -\frac{1}{8\pi^2}\, I_{12}^2\left( \ln \frac{\epsilon^2}{x_{12}^2} -3 \right)\,.
\label{pinchinglimits}
\end{aligned}
\ee
The strategy for the computation of one-loop corrections to
higher point functions of chiral primary operators
$\cO^{u_i}_k(x)=\Tr[(u^I_i\, \phi^I(x))^k]$ 
will be to consider the ``dressing'' of the tree-level graphs. We shall
distinguish the following interaction insertions types:

\textbf{(i) Corner interactions.} Combining one-half of the self-energy corrections
on every leg with the gluon exchange and four-point interaction one has for every corner
\be
\label{corner}
\quad\ \,\raisebox{0mm}[9mm][9mm]{\parbox{8mm}{
\begin{fmfgraph*}(6,9) 
        \fmfstraight
        \fmfbottom{w2,u2}\fmfleft{u1}\fmftop{w3,u3}
        \fmf{plain,width=1}{u1,v1,u2}
        \fmf{plain,width=1}{u1,v2,u3}
        \fmffreeze
        \fmf{dashes,width=1}{v1,v2}
        \fmfblob{.4w}{u1}
        \fmflabel{$u_2$}{u1}
        \fmflabel{$u_1$}{u2}
        \fmflabel{$u_3$}{u3}
        \fmfdot{u1,u2,u3}
\end{fmfgraph*}}} 
:= \frac{1}{2}\left [\ 
\raisebox{0mm}[5mm][5mm]{\parbox{7mm}{
\begin{fmfgraph}(6,9) 
        \fmfstraight
        \fmfbottom{w2,u2}\fmfleft{u1}\fmftop{w3,u3}
        \fmf{plain,width=1}{u1,v1,u2}
        \fmf{plain,width=1}{u1,v2,u3}
        \fmffreeze
        \fmfblob{.4w}{v1}
        \fmfdot{u1,u2,u3}
\end{fmfgraph}}} \,\,
+ \ 
\raisebox{0mm}[5mm][5mm]{\parbox{8mm}{
\begin{fmfgraph}(6,9) 
        \fmfstraight
        \fmfbottom{w2,u2}\fmfleft{u1}\fmftop{w3,u3}
        \fmf{plain,width=1}{u1,v1,u2}
        \fmf{plain,width=1}{u1,v2,u3}
        \fmffreeze
        \fmfblob{.4w}{v2}
        \fmfdot{u1,u2,u3}
\end{fmfgraph}}}
\right ] 
+ \,\,
\raisebox{0mm}[6mm][6mm]{\parbox{7mm}{
\begin{fmfgraph}(6,9) 
        \fmfstraight
        \fmfbottom{w2,u2}\fmfleft{u1}\fmftop{w3,u3}
        \fmf{plain,width=1}{u1,v1,u2}
        \fmf{plain,width=1}{u1,v2,u3}
        \fmffreeze
        \fmf{wiggly}{v1,v2}
        \fmfdot{u1,u2,u3}
\end{fmfgraph}}}
+ \, \,
\raisebox{0mm}[6mm][6mm]{\parbox{7mm}{
\begin{fmfgraph}(6,9) 
        \fmfstraight
        \fmfbottom{w2,u2}\fmfleft{u1}\fmftop{w3,u3}
       \fmf{plain,width=1}{u2,v,u3}
        \fmf{plain,left,width=1}{u1,v,u1} 
       \fmfdot{u1,u2,u3}
\end{fmfgraph}}}
= C_{123}[12] [23]\,.
\ee
where we have defined the corner insertion
\be
C_{123} :=\frac{\lambda}{2} \, Y_{123}\left(\frac{1}{I_{12}} +\frac{1}{I_{23}} 
-\frac{2}{I_{31}} \right).
\ee
Using this result and the pinching identities \eqn{pinchinglimits} 
it is easy to check the vanishing
of the one-loop correction to the two-point functions
$\langle \,\cO_2(x_1)\,\cO_2(x_2) \,\rangle$
\be
\left.\raisebox{0mm}[5mm][5mm]{\parbox{7mm}{
\begin{fmfgraph*}(5,5) 
        \fmftop{u1}
        \fmfbottom{u2}
        \fmfdot{u1,u2}
        \fmflabel{$u_1$}{u1}
        \fmflabel{$u_2$}{u2}
        \fmf{plain,left,width=1}{u1,u2,u1}
\end{fmfgraph*}}}
\right|_{\textstyle\genfrac{}{}{0pt}{}{\text{1-loop}}{\text{dressed}}}
= 2^2\, ( C_{212}+C_{121} )\, [12]^2= 0 \,.
\ee
Similarly one shows the vanishing of the one-loop correction to the 
three-point functions
$\langle \,\cO_2(x_1)\,\cO_2(x_2) \,\cO_2(x_3)\,\rangle$
\be
\label{3-pt}
\left.\raisebox{0mm}[8mm][6mm]{\parbox{13mm}{
\begin{fmfgraph*}(6,8) 
        \fmfstraight
        \fmftop{u1,w1}
        \fmfright{u2}
        \fmfbottom{u3,w3}
        \fmfdot{u1,u2,u3}
        \fmflabel{$u_1$}{u1}
        \fmflabel{$u_2$}{u2}
        \fmflabel{$u_3$}{u3}
        \fmf{plain,width=1}{u1,u2,u3,u1}
        \end{fmfgraph*}}}
\right|_{\textstyle\genfrac{}{}{0pt}{}{\text{1-loop}}{\text{dressed}}}
= 2^3\, ( C_{123}+C_{231}+C_{312} )
\, [12][23][31]= 0 \,.
\ee

\textbf{(ii) Non-corner interactions.} Here we have to consider
 the gluon exchange and scalar four-point
graphs of \eqn{Ginsertion} and \eqn{Xinsertion}. These graphs involve the 
invariant cross-ratios and the scalar graph also includes a contraction which 
at tree-level would not be planar: The crossed contraction $[14][23]$ in the 
numbering convention of \eqn{Xinsertion}. From now on, however, we shall 
switch to a cyclic ordering of the points $\{1,2,3,4\}$ and define
\be
\label{4pt-scalar}
S_{1234} :=
\quad\ \,
\raisebox{0mm}[8mm][2mm]{\parbox{14mm}{
\begin{fmfgraph*}(8,8) \fmfstraight
        \fmftop{u1,u2}
        \fmfbottom{u3,u4}
        \fmflabel{$u_2$}{u2}
       \fmflabel{$u_3$}{u4}
        \fmflabel{$u_1$}{u1}
        \fmflabel{$u_4$}{u3}
       \fmfdot{u1,u2,u3,u4}
        \fmf{plain,width=1}{u1,v,u4}
        \fmf{plain,width=1}{u3,v,u2}
\end{fmfgraph*}}}
=\frac{\lambda}{2}\,\frac{X_{1234}}{I_{13}I_{24}}
\big(\, 2 \,[13][24]-t\, [23][14]-s\,[12][34]\big)\,,\\[0.4cm]
\ee
where we have used the cross-ratios 
$s=\frac{I_{13}I_{24}}{I_{12}I_{34}}$ and
$t=\frac{I_{13}I_{24}}{I_{14}I_{23}}$
of \eqn{cross-ratios}.

For the gluon exchange graph \eqn{Ginsertion} we define
\begin{align}
\label{4pt-gluon}
G_{1234}&:=
\quad\ \,
\raisebox{0mm}[8mm][8mm]
{\parbox{14mm}{
\begin{fmfgraph*}(8,8) \fmfstraight
        \fmftop{u1,u2}
        \fmfbottom{u4,u3}
        \fmfdot{u1,u2,u3,u4}
        \fmflabel{$u_2$}{u2}
        \fmflabel{$u_4$}{u4}
        \fmflabel{$u_1$}{u1}
        \fmflabel{$u_3$}{u3}
        \fmf{plain,width=1}{u1,v1,u2}
        \fmf{plain,width=1}{u3,v2,u4}\fmffreeze
        \fmf{wiggly}{v1,v2}
\end{fmfgraph*}}}
=\frac{\lambda}{2}\,F_{12,43}\, [12][34]
=\,\bigg[\frac{\lambda}{2}\,\frac{X_{1234}}{I_{13}I_{24}}\big(t-1\big)+C_{12,43}\bigg][12][34]
\\
C_{12,43}&:=\frac{1}{3}\big(C_{123}+C_{412}+C_{341}+C_{234}
-C_{124}-C_{243}-C_{431}-C_{312}\big)\,.
\label{C4def}
\end{align}
Note that this expression does preserve the structure of the tree-level contractions. 
In the final expression in \eqn{4pt-gluon} we wrote $F_{12,43}$ as a sum of two 
terms, the first involving the box integral $\Phi(s,t)$ \eqn{Phi} and the second 
which is a sum of terms identical to the corner graphs \eqn{corner} and which 
does not involve the cross-ratios.

Lastly we define a full four-point planar interaction insertion by combining both orientations and removing the corner interactions
\begin{align}
D_{1234}&:=
\quad\ \,
\raisebox{0mm}[8mm][8mm]{\parbox{14mm}{
\begin{fmfgraph*}(8,8) \fmfstraight
        \fmftop{u1,u2}
        \fmfbottom{u4,u3}
        \fmfdot{u1,u2,u3,u4}
        \fmflabel{$u_2$}{u2}
        \fmflabel{$u_4$}{u4}
        \fmflabel{$u_1$}{u1}
        \fmflabel{$u_3$}{u3}
        \fmf{plain,width=1}{u1,v,u2}
        \fmf{plain,width=1}{u3,v,u4}\fmffreeze
        \fmfblob{.4w}{v}
\end{fmfgraph*}}}
=\ \, 
\parbox{10mm}{
\begin{fmfgraph}(8,8) \fmfstraight
        \fmftop{u1,u2}
        \fmfbottom{u4,u3}
        \fmfdot{u1,u2,u3,u4}
        \fmf{plain,width=1}{u1,v1,u2}
        \fmf{plain,width=1}{u3,v2,u4}
        \fmffreeze
        \fmf{wiggly}{v1,v2}
\end{fmfgraph}}
+\ \,
\parbox{10mm}{
\begin{fmfgraph}(8,8) \fmfstraight
        \fmftop{u1,u2}
        \fmfbottom{u3,u4}
        \fmfdot{u1,u2,u3,u4}
        \fmf{plain,width=1}{u1,v1,u3}
        \fmf{plain,width=1}{u2,v2,u4}
        \fmffreeze
        \fmf{wiggly}{v1,v2}
\end{fmfgraph}}
+\ \,
\parbox{10mm}{
\begin{fmfgraph}(8,8) \fmfstraight
        \fmftop{u1,u2}
        \fmfbottom{u4,u3}
        \fmfdot{u1,u2,u3,u4}
        \fmf{plain,width=1}{u1,v,u4}
        \fmf{plain,width=1}{u3,v,u2}
\end{fmfgraph}}
-C_{12,43}[12][34]-C_{14,23}[14][23]
\nn\\&=
\frac{\lambda}{2}\,\frac{X_{1234}}{I_{13}I_{24}}
\big(2\,[13][24]+(s-1-t)[14][23]+(t-1-s)\,[12][34]\big) \nn\\
&= \frac{\lambda}{32\pi^2}\, \Phi(s,t)\,
\big(2\,[13][24]+(s-1-t)[14][23]+(t-1-s)\,[12][34]\big) 
\label{D-graph}
\end{align}
Note that $D_{1234}$ has manifest cyclic symmetry as well as reflection symmetry $2\leftrightarrow4$ (which is accompanied, of course, by $s\leftrightarrow t$).

\subsection{Cancelation of corner graphs}

We would like to calculate the full connected one-loop planar interacting diagrams. In order to enumerate these
we can relate them to tree-level diagrams by cutting the interacting line. Cutting the dashed line in the
combined corner interaction \eqn{corner} gives a single tree-level graph. The same is true for the four-point gluon exchange \eqn{4pt-gluon}, but not for the four-scalar vertex \eqn{4pt-scalar} as it can be cut in two
different ways. One point to note, though, is that the underlying tree-level graph of a 
non-corner interaction may be disconnected.

This allows us therefore to go in the opposite direction, starting with all tree-level planar graphs (including disconnected ones) and dress them up with one-loop interactions. This dressing involves the exchange of gluons (and also the scalar interaction) between pairs of lines which form the border of a face on the graph.

Consider such a face with $n$ vertices at positions $x_1,\ldots x_n$ cyclically ordered
and with the local operators at $x_i$ formulated by contracting the scalar fields with the six-vectors 
$u_i^I$. At one-loop there will be the following corner interactions within this face
\be
\sum_{i=1}^nC_{(i-1)\,i\,(i+1)}\,,\qquad i_0\equiv i_n\,.
\ee
In addition there will be four-point interactions from all pairs of non-adjacent links, where we strip away the tree-level contractions. Let us now consider the sum of the corner interactions and the 
corner-like contributions $C_{ij,lm}$ to the four-point interactions 
\begin{align}
\sum_{i=1}^n \big (C_{i-1\,i\,i+1}
+\frac{1}{2}\sum_{j\neq i-1,i,i+1}C_{i\,i+1,j\,j+1} \, \big )
&=\sum_{i=1}^n\bigg(C_{i-1\,i\,i+1}
-\frac{1}{6}\sum_{j=i+2}^{i-2}\big(C_{i\,i+1\,j}+C_{i+1\,j\,j+1}
\nn\\&\hskip-5.5cm
+C_{j\,j+1\,i}+C_{j+1\,i\,i+1}
-C_{i\,i+1\,j+1}-C_{i+1\,i\,j}-C_{j\,j+1\,i+1}-C_{j+1\,j\,i}\big)\bigg)=0\,.
\end{align}
To see this cancelation it is convenient to rearrange the sum by the differences of indices, noting that $C_{ijl}=C_{lji}$. Lastly one needs the relation $C_{ijl}+C_{jli}+C_{lij}=0$ \eqn{3-pt}.

\subsection{General one-loop insertion formula}

Having proved that all corner interactions cancel we can write down the 
general one-loop four-point insertion formula. Starting with any planar 
tree graph we dress up every pair of links that are on one face but not 
adjacent with the basic four-point insertion $\frac{1}{2}D_{ijlm}$ \eqn{D-graph}. 
The factor of $1/2$ is there because the same insertion will come from 
another tree graph with the contraction $[ij][lm]$ replaced by $[jl][mi]$.

We get the general one-loop result
\be
\label{4-pt}
\vev{\cO^{u_1}_{k_1}\cdots\cO^{u_n}_{k_n}}_\text{1-loop}
=\sum_{i,j,l,m}k_ik_jk_lk_m\,D_{ijlm} 
\vev{\cO^{u_i}_{k_i-1}\cO^{u_j}_{k_j-1}\cO^{u_l}_{k_l-1}\cO^{u_m}_{k_m-1}\bigg|
\prod_{p\neq i,j,l,m}\!\cO^{u_p}_{k_p}}_\text{tree, disc}
\ee
The expression on the right-hand side is the tree level $n$-point function of 
the operators where one scalar is removed from each of the four operators 
and this should be a planar diagram with the topology of a disk, with the four 
operators with labels $ijlm$ inserted (in this order) at the boundary of the disc 
and the rest in the bulk. The contractions represented by $D_{ijlm}$ are then 
outside of the disc, but still have a planar spherical topology. This amplitude 
is represented pictorially in Figure~\ref{fig-disc}.

\begin{figure}
\begin{center}
$$
\vev{\cO^{u_i}_{k_i-1}\cO^{u_j}_{k_j-1}\cO^{u_l}_{k_l-1}\cO^{u_m}_{k_m-1}\bigg|
\prod_{p\neq i,j,l,m}\!\!\cO^{u_p}_{k_p}}_\text{tree, disc}
\!\!\!\!=\qquad\quad\  
\raisebox{0mm}[15mm][12mm]{\parbox{30mm}{
\begin{fmfgraph*}(20,20) \fmfstraight
        \fmfsurround{u1,u2,u3,u4}
        \fmfdot{u1,u2,u3,u4}
        \fmflabel{$\cO^{u_i}_{k_i-1}$}{u2}
        \fmflabel{$\cO^{u_j}_{k_j-1}$}{u3}
        \fmflabel{$\cO^{u_l}_{k_l-1}$}{u4}
        \fmflabel{$\cO^{u_m}_{k_m-1}$}{u1}
        \fmffreeze
        \fmf{phantom}{u1,v,u3}
        \fmf{phantom}{u1,w1,u2,w2,u3,w3,u4,w4,u1}
        \fmf{phantom}{w1,w5,w2,u4}
        \fmf{phantom}{w2,w5,w1}
        \fmf{phantom}{u2,w6,w7,u4}
        \fmfdot{w1,w2,w4,w7}
        \fmfv{label=$\cO^{u_p}_{k_p}$,label.angle=0,label.dist=-.2}{v}
        \fmf{dashes,right=.4}{u1,u2,u3,u4,u1}
\end{fmfgraph*}}}
$$
\parbox{13cm}{
\caption{Graphical representation of the disc correlation function of 
equation~\eqn{4-pt}. The first four operators are placed on the boundary 
of the disc and all the others are in the interior. In \eqn{4-pt} one is 
instructed to sum over all planar tree-level contractions of these operators.
\label{fig-disc}}}
\end{center}
\end{figure}

$D_{ijlm}$ \eqn{D-graph} involves a transcendental function $\Phi(s,t)$ \eqn{Phi} 
of the cross-ratios of the four points $x_i$, $x_j$, $x_l$ and $x_m$. Since the 
tree-level contraction among these four operators and all the others involves 
only rational functions, it is natural to separate the graphs in this way. When looking 
for special cancelations in the one-loop amplitudes at generic positions (as done 
in \cite{us2}) there are no algebraic relations among the $D$-functions. The 
exception are terms with the same boundary vertices $i$, $j$, $k$ and $l$, but 
in a different order, since the functions $D_{ijlm}$ are then related to each-other. 
Due to the cyclic and reflection symmetry of $D_{ijlm}$, there are 
only three inequivalent orderings $ijlm$, $iljm$ and $ijml$.

The crucial relation we use extensively is
\be
D_{ijlm}+D_{iljm}+D_{ijml}=0\,.
\label{modular}
\ee

\section{Four-point functions}
\label{sec-4-pt}

We would like here to rederive the factorization formula of the four-point 
function \eqn{eins} of \cite{Arutyunov:2002fh} at one-loop using the 
general formula \eqn{4-pt}. It is instructive to start with the simplest 
four-point function, of four operators of dimension two.

\subsection{Four-point functions of $\cO_2^{u_i}$}

Applying our general formula \eqn{4-pt} in this case we have
\be
\begin{aligned}
\vev{\cO^{u_1}_2\cO^{u_2}_2\cO^{u_3}_2\cO^{u_4}_2}_\text{1-loop}
=&\,16\big(D_{1234}\,\vev{\cO^{u_1}_1\cO^{u_2}_1\cO^{u_3}_1\cO^{u_4}_1}_\text{tree, disc}
\\&\hskip-2cm
+D_{1324}\,\vev{\cO^{u_1}_1\cO^{u_3}_1\cO^{u_2}_1\cO^{u_4}_1}_\text{tree, disc}
+D_{1243}\,\vev{\cO^{u_1}_1\cO^{u_2}_1\cO^{u_4}_1\cO^{u_3}_1}_\text{tree, disc}\big).
\label{V1111}
\end{aligned}
\ee
For a given ordering there are two planar tree diagrams on the disc, with a pair of 
contractions labeled $\cX$, $\cY$ and $\cZ$ of \eqn{XYZ}. We get
\be
\begin{aligned}
\vev{\cO^{u_1}_2\cO^{u_2}_2\cO^{u_3}_2\cO^{u_4}_2}_\text{1-loop}
=&\,16\big(D_{1234}(\cX+\cZ)
+D_{1243}(\cY+\cX)
+D_{1324}(\cZ+\cY)\big)
\\
=&\,-\,16\big(D_{1234}\cY+D_{1243}\cZ+D_{1324}\cX\big).
\end{aligned}
\label{4pt-2222-temp}
\ee
To get the last line we subtracted from all the terms the sum of all three pair-wise 
contractions $\cX+\cY+\cZ$ and used the fact that 
$(D_{1234}+D_{1243}+D_{1324})=0$ \eqn{modular}. 
So the result is written as minus the sum of all {\em non}-planar contractions.

Now we note that we can also express $D_{1234}$ of \eqn{D-graph} in terms 
of the pair-wise contractions and the universal polynomial 
prefactor $\cRNf$ of \eqn{cRN4} as
\be
D_{1234}
=
\frac{\lambda}{2}\,\frac{X_{1234}}{I_{13}I_{24}}
\big(2\cY-\cX-\cZ+(s-t)(\cZ-\cX)\big)
=\frac{\lambda}{2}\,\frac{\Phi(s,t)}{16\pi^2}
\,\frac{\partial \cRNf}{\partial\cY}\,,
\ee
and likewise for the other ones. This gives
\be
\begin{aligned}
\vev{\cO^{u_1}_2\cO^{u_2}_2\cO^{u_3}_2\cO^{u_4}_2}_\text{1-loop}
&=-\frac{\lambda}{2\pi^2}\,\Phi(s,t)
\left(\cY\frac{\partial \cRNf}{\partial\cY}
+\cZ\frac{\partial \cRNf}{\partial\cZ}
+\cX\frac{\partial \cRNf}{\partial\cX}\right)
\\&
=-\frac{\lambda}{\pi^2}\,\Phi(s,t)\,\cRNf
\end{aligned}
\label{4pt-2222}
\ee

\subsection{General four-point function}

Now consider a general four-point function
\be
\vev{\cO^{u_1}_{k_1}\cO^{u_2}_{k_2}\cO^{u_3}_{k_3}\cO^{u_4}_{k_4}}\,.
\ee
Applying our general formula \eqn{4-pt} we have
\be
\begin{aligned}
\vev{\cO^{u_1}_{k_1}\cO^{u_2}_{k_2}\cO^{u_3}_{k_3}\cO^{u_4}_{k_4}}_\text{1-loop}
=k_1k_2k_3k_4\Big(&D_{1234}\,
\vev{\cO^{u_1}_{k_1-1}\cO^{u_2}_{k_2-1}
\cO^{u_3}_{k_3-1}\cO^{u_4}_{k_4-1}}_\text{tree, disc}
\\&+
D_{1324}\,\vev{\cO^{u_1}_{k_1-1}\cO^{u_3}_{k_3-1}\cO^{u_2}_{k_2-1}
\cO^{u_4}_{k_4-1}}_\text{tree, disc}
\\&+
D_{1243}\,\vev{\cO^{u_1}_{k_1-1}\cO^{u_2}_{k_2-1}\cO^{u_4}_{k_4-1}
\cO^{u_3}_{k_3-1}}_\text{tree, disc}\Big).
\end{aligned}
\label{general-4pt}
\ee
Let us examine the term multiplying $D_{1234}$. It involves all possible tree-level 
planar contractions on the disc. It is clear that it can be factorized as
\be
\vev{\cO^{u_1}_{k_1-1}\cO^{u_2}_{k_2-1}
\cO^{u_3}_{k_3-1}\cO^{u_4}_{k_4-1}}_\text{tree, disc}
=\big([12][34]+[14][23]\big)
\times\big\{\cdots\big\}
\ee
where $\{\cdots\}$ stands for some planar tree-level contractions of
$\cO^{u_1}_{k_1-2}\cO^{u_2}_{k_2-2}
\cO^{u_3}_{k_3-2}\cO^{u_4}_{k_4-2}$.

The non-trivial fact about the factorization formula of Arutyunov et al. 
\cite{Arutyunov:2002fh,Arutyunov:2003ae}, is that exactly the same combinatorics 
of tree-contractions appear in all three permutations of the order of insertions 
in \eqn{general-4pt}. Writing the pairwise contractions as $\cX$, $\cY$ and $\cZ$ 
\eqn{XYZ} the four point function becomes
\begin{align}
\vev{\cO^{u_1}_{k_1}\cO^{u_2}_{k_2}\cO^{u_3}_{k_3}\cO^{u_4}_{k_4}}_\text{1-loop}
\\&\hskip-2cm
=k_1k_2k_3k_4
\big(D_{1234}(\cX+\cZ)+D_{1243}(\cY+\cX)+D_{1324}(\cZ+\cY)\big)
\big\{\cdots\big\}
\nn\\&\hskip-2cm
=-\frac{k_1k_2k_3k_4}{16}\frac{\lambda}{\pi^2}\,\Phi(s,t)\,\cRNf\,
\big\{\cdots\big\}\,.
\end{align}
Indeed when $k_1=k_2=k_3=k_4$, a simple inspection reveals that the same 
combinatorics appear in all the terms in \eqn{general-4pt}.

\section{Five-point functions}
\label{sec-5-pt}

\subsection{Five-point functions of $\cO_2^{u_i}$}

We turn now to the simplest five-point function, that of five operators of 
dimension two. Applying our general formula \eqn{4-pt} in this case we have
\begin{align}
\label{5pt-22222-0}
\vev{\cO^{u_1}_2\cO^{u_2}_2\cO^{u_3}_2\cO^{u_4}_2\cO^{u_5}_2}_\text{1-loop}
&=16\,\big ( D_{1234}\, \vev{\cO^{u_1}_1\cO^{u_2}_1\cO^{u_3}_1\cO^{u_4}_1|\cO^{u_5}_2}_\text{tree, disc}
\\&\hskip-3cm
+D_{1324}\,\vev{\cO^{u_1}_1\cO^{u_3}_1\cO^{u_2}_1\cO^{u_4}_1|\cO^{u_5}_2}_\text{tree, disc}
+D_{1243}\,\vev{\cO^{u_1}_1\cO^{u_2}_1\cO^{u_4}_1\cO^{u_3}_1|\cO^{u_5}_2}_\text{tree, disc} \big)
\nn\\&\hskip-3.5cm
+16\, \big ( D_{1235}\, \vev{\cO^{u_1}_1\cO^{u_2}_1\cO^{u_3}_1\cO^{u_5}_1|\cO^{u_4}_2}_\text{tree, disc}+
\cdots\big )
\nn\\&\hskip-3.5cm
+16\, \big (  D_{1245}\,  \vev{\cO^{u_1}_1\cO^{u_2}_1\cO^{u_4}_1\cO^{u_5}_1|\cO^{u_3}_2}_\text{tree, disc} +\cdots\big ) 
\nn\\&\hskip-3.5cm
+16\, \big ( D_{1345}\,  \vev{\cO^{u_1}_1\cO^{u_3}_1\cO^{u_4}_1\cO^{u_5}_1|\cO^{u_2}_2}_\text{tree, disc}+ \cdots \big )
\nn\\&\hskip-3.5cm
+16\, \big ( D_{2345}\,  \vev{\cO^{u_2}_1\cO^{u_3}_1\cO^{u_4}_1\cO^{u_5}_1|\cO^{u_1}_2}_\text{tree, disc} + \cdots \big)\,.
\nn
\end{align}
The function $D_{ijlm}$ is given in \eqn{D-graph}. To evaluate 
\eqn{5pt-22222-0} we need to find the 
tree level disc amplitudes, which are all the same, up to permutations of indices. 
A simple application of Wick's theorem gives
\be
\vev{\cO^{u_1}_1\cO^{u_2}_1\cO^{u_3}_1\cO^{u_4}_1|\cO^{u_5}_2}_\text{tree, disc}
=2\,\big([12,34|5]+[14,23|5]\big)\,
\label{V11112}
\ee
where we define the pair-wise contraction through a fifth point as 
a generalization of $\cX$, $\cY$, $\cZ$ \eqn{XYZ}
\be
[12,34|5]=[15][25][34]+[12][35][45]\,.
\ee

As in the case of the four-point function in Section~\ref{sec-4-pt}, we subtract 
the sum of all three possible contractions 
$([12,34|5]+[13,24|5]+[14,23|5])$ from the first three terms of 
\eqn{5pt-22222-0} using the fact that 
$(D_{1234}+D_{1243}+D_{1324})=0$. After similar manipulations of 
all the other terms, this allows us to express \eqn{5pt-22222-0} as minus 
the sum over all non-planar graphs
\be
\begin{aligned}
\label{5pt-22222}
\vev{\cO^{u_1}_2\cdots
\cO^{u_5}_2}_\text{1-loop}
=&-32\,\big ( D_{1234}[13,24|5]+D_{1324}[12,34|5]+D_{1243}[14,23|5]\big)
\\&
-32\,\big ( D_{1235}[13,25|4]+D_{1325}[12,53|4]+D_{1253}[15,23|4]\big)
\\&
-32\,\big ( D_{1254}[15,24|3]+D_{1524}[12,45|3]+D_{1245}[14,25|3]\big)
\\&
-32\,\big ( D_{1534}[13,54|2]+D_{1354}[15,34|2]+D_{1543}[14,53|2]\big)
\\&
-32\,\big ( D_{5234}[53,24|1]+D_{5324}[52,34|1]+D_{5243}[54,23|1]\big)
\end{aligned}
\ee

\subsection{Five-point functions of four $\cO_2^{u_i}$ and one $\cO_4^{u_5}$}

Next we look at the five-point function with four operators of 
dimension two and one of dimension four. 
Applying our general formula \eqn{4-pt} in this case we have
\begin{align}
\label{5pt-22224}
\vev{\cO^{u_1}_2\cO^{u_2}_2\cO^{u_3}_2\cO^{u_4}_2\cO^{u_5}_4}_\text{1-loop}
&=16\, \big( D_{1234}\,
\vev{\cO^{u_1}_1\cO^{u_2}_1\cO^{u_3}_1\cO^{u_4}_1|\cO^{u_5}_4}_\text{tree, disc}
\\&\hskip-3cm
+D_{1324}\,\vev{\cO^{u_1}_1\cO^{u_3}_1\cO^{u_2}_1\cO^{u_4}_1|\cO^{u_5}_4}_\text{tree, disc}
+D_{1243}\,\vev{\cO^{u_1}_1\cO^{u_2}_1\cO^{u_4}_1\cO^{u_3}_1|\cO^{u_5}_4}_\text{tree, disc}\big)
\nn\\&\hskip-3cm
+32\, \big( D_{1235} \, \vev{\cO^{u_1}_1\cO^{u_2}_1\cO^{u_3}_1\cO^{u_5}_3|\cO^{u_4}_2}_\text{tree, disc}+\cdots\big)
\nn\\&\hskip-3cm
+32\, \big( D_{1245}\,\vev{\cO^{u_1}_1\cO^{u_2}_1\cO^{u_4}_1\cO^{u_5}_3|\cO^{u_3}_2}_\text{tree, disc} +\cdots\big)
\nn\\&\hskip-3cm
+32\, \big( D_{1345}\, \vev{\cO^{u_1}_1\cO^{u_3}_1\cO^{u_4}_1\cO^{u_5}_3|\cO^{u_2}_2}_\text{tree, disc}+\cdots\big)
\nn\\&\hskip-3cm
+32\, \big(D_{2345}\, \vev{\cO^{u_2}_1\cO^{u_3}_1\cO^{u_4}_1\cO^{u_5}_3|\cO^{u_1}_2}_\text{tree, disc}+ \cdots\big)\,.\nn
\end{align}
The first tree-level disc amplitude in this expression is
\be
\vev{\cO^{u_1}_1\cO^{u_2}_1\cO^{u_3}_1\cO^{u_4}_1|\cO^{u_5}_4}_\text{tree, disc}
=4\,[15][25][35][45]\,.
\ee
this is clearly independent of the order of the labels 1234, so the sum of 
the first three terms in \eqn{5pt-22224} is proportional to
$(D_{1234}+D_{1324}+D_{1243})$, which vanishes by \eqn{modular}.

The other $D_{ijlm}$ multiply other tree-level amplitudes. For example 
$D_{1235}$ multiplies the term
\begin{align}
\vev{\cO^{u_1}_1\cO^{u_2}_1\cO^{u_3}_1\cO^{u_5}_3|\cO^{u_4}_2}_\text{tree, disc}
=2\,\Big( &[35][15][45][24]+[25][35][45][14]+[25][15][45][34]
\nn\\&
+2[45]^2\big([35][12]+[15][23]\big)\Big).
\end{align}
Now we note that under permutations of 1, 2 and 3 the terms in the first line get 
interchanged. Therefore these will multiply the sum $(D_{1235}+D_{1325}+D_{1253})$ 
and hence vanish (recall that $D_{1253}=D_{2135}$). We are left with the 
terms on the second line, which may be written as $2[45]^2\,
\vev{\cO^{u_1}_1\cO^{u_2}_1\cO^{u_3}_1\cO^{u_5}_1}_\text{tree, disc}$, 
and the sum over the three permutations is proportional to the one-loop 
four-point function
\be
[45]^2\vev{\cO^{u_1}_2\cO^{u_2}_2\cO^{u_3}_2\cO^{u_5}_2}_\text{1-loop}
\ee
The five-point function \eqn{5pt-22224} is therefore a sum over four terms, 
each proportional to a one-loop four-point function
\begin{align}
\!\!\vev{\cO^{u_1}_2\cO^{u_2}_2\cO^{u_3}_2\cO^{u_4}_2\cO^{u_5}_4}_\text{1-loop}
=&\,
8[15]^2\vev{\cO^{u_2}_2\cO^{u_3}_2\cO^{u_4}_2\cO^{u_5}_2}_\text{1-loop}
+8[25]^2\vev{\cO^{u_1}_2\cO^{u_3}_2\cO^{u_4}_2\cO^{u_5}_2}_\text{1-loop}
\nn\\&\hskip-2cm
+8[35]^2\vev{\cO^{u_1}_2\cO^{u_2}_2\cO^{u_4}_2\cO^{u_5}_2}_\text{1-loop}\
+8[45]^2\vev{\cO^{u_1}_2\cO^{u_2}_2\cO^{u_3}_2\cO^{u_5}_2}_\text{1-loop}\,.
\end{align}
The reduction of this near-extremal correlator to weight two four-point function
was also observed in \cite{nearext}.

\subsection{Five-point functions of three $\cO_2^{u_i}$ and two $\cO_3^{u_i}$}

The next five-point function we evaluate is
\be
\label{5pt-22233}
\vev{\cO^{u_1}_2\cO^{u_2}_2\cO^{u_3}_2\cO^{u_4}_3\cO^{u_5}_3}_\text{1-loop}
\ee
Like in the last example we have to consider a few inequivalent disc amplitudes 
to calculate \eqn{4-pt}. First there are the amplitudes multiplying $D_{1234}$ 
and $D_{1235}$ with different orderings which are of the form
\be
\vev{\cO^{u_1}_1\cO^{u_2}_1\cO^{u_3}_1\cO^{u_4}_2|\cO^{u_5}_3}_\text{tree, disc}
=3 \, [45]\big([12,34|5]+[14,23|5]\big)\,.
\label{V11123}
\ee
These are proportional to the terms appearing in the one-loop correction to the 
simplest five-point function \eqn{V11112}.

Then, when the interaction is among the points 1245, there are two 
different inequivalent orderings of the 
vertices, when 4 and 5 are adjacent or not. In the first case we have
\be
\begin{aligned}
\vev{\cO^{u_1}_1\cO^{u_2}_1\cO^{u_4}_2\cO^{u_5}_2|\cO^{u_3}_2}_\text{tree, disc}
=&\,2\Big([14][24][35]^2+[15][25][34]^2
\\&\hskip-2cm
+[34][35]\big([12][45]+[15][24]\big)+[45]\big([12,45|3]+[15,24|3]\big)\Big)
\end{aligned}
\label{V11222}
\ee
and likewise permuting $1\leftrightarrow2$
\be
\begin{aligned}
\vev{\cO^{u_2}_1\cO^{u_1}_1\cO^{u_4}_2\cO^{u_5}_2|\cO^{u_3}_2}_\text{tree, disc}
=&\,2\Big([14][24][35]^2+[15][25][34]^2
\\&\hskip-2cm
+[34][35]\big([12][45]+[25][14]\big)+[45]\big([12,45|3]+[14,25|3]\big)\Big)
\end{aligned}
\ee
For the third, the inequivalent, ordering we have
\be
\begin{aligned}
\vev{\cO^{u_1}_1\cO^{u_4}_2\cO^{u_2}_1\cO^{u_5}_2|\cO^{u_3}_2}_\text{tree, disc}
=&\,2\Big([14][24][35]^2+[15][25][34]^2
\\&\hskip-2cm
+[34][35]\big([14][25]+[15][24]\big)+[45]\big[14,25|3]+[15,24|3]\big)\Big).
\end{aligned}
\label{V12122}
\ee
Now we note that $[14][24][35]^2$ and $[15][25][34]^2$ appear in each of the 
three permutations, and therefore vanish.

As before, by subtracting the sum of the contractions appearing above we get 
the simple expression for the terms multiplying $D_{1245}$ and its permutations
\be
\begin{aligned}
&-[34][35]\big(D_{1245}[14][25]+D_{2145}[15][24]+D_{1425}[12][45]\big)
\\&-[45]\big(D_{1245}[14,25|3]+D_{2145}[15,24|3]+D_{1425}[12,45|3]\big)
\end{aligned}
\ee
Note that the three terms on the first line are proportional to the 
one-loop four-point function of operators of dimension two \eqn{V1111}. 
The terms on the second line are like those appearing in the expression for 
the simplest five-point function \eqn{5pt-22222}.

Summing all the different terms gives the one-loop five-point amplitude as 
a sum over the simplest four-point function and the simplest five-point function
\be
\begin{aligned}
&\vev{\cO^{u_1}_2\cO^{u_2}_2\cO^{u_3}_2\cO^{u_4}_3\cO^{u_5}_3}_\text{1-loop}
=
\frac{9}{4}\Big(
[45]\vev{\cO^{u_1}_2\cO^{u_2}_2\cO^{u_3}_2\cO^{u_4}_2\cO^{u_5}_2}_\text{1-loop}
\\&\hskip2cm
+2\,[41][15]\vev{\cO^{u_2}_2\cO^{u_3}_2\cO^{u_4}_2\cO^{u_5}_2}_\text{1-loop}
+2\,[42][25]\vev{\cO^{u_1}_2\cO^{u_3}_2\cO^{u_4}_2\cO^{u_5}_2}_\text{1-loop}
\\&\hskip2cm
+2\,[43][35]\vev{\cO^{u_1}_2\cO^{u_2}_2\cO^{u_4}_2\cO^{u_5}_2}_\text{1-loop}\Big).
\end{aligned}
\ee

\subsection{General five-point function}

We do not have a general explicit formula for the one-loop five-point function, 
but beyond the three examples above we calculated eleven more 
examples of five-point functions with operators of total dimension up to 
sixteen. They are listed in Appendix~\ref{app}.

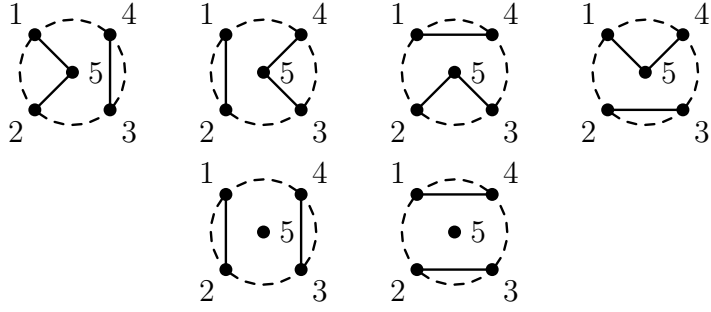
\begin{figure}
\begin{center}
$$
\begin{gathered}
\qquad\quad
\raisebox{0mm}[8mm][10mm]{\parbox{13mm}{
\begin{fmfgraph*}(10,10) \fmfstraight
        \fmftop{u1,u4}
        \fmfbottom{u2,u3}
        \fmf{phantom}{u1,u5,u3}
        \fmfdot{u1,u2,u3,u4,u5}
        \fmflabel{1}{u1}
        \fmflabel{2}{u2}
        \fmflabel{3}{u3}
        \fmflabel{4}{u4}
        \fmflabel{5}{u5}
        \fmffreeze
        \fmf{dashes,right=.4}{u1,u2,u3,u4,u1}
        \fmf{plain}{u1,u5,u2}
        \fmf{plain}{u3,u4}
\end{fmfgraph*}}}
\qquad\quad
\raisebox{0mm}[8mm][10mm]{\parbox{13mm}{
\begin{fmfgraph*}(10,10) \fmfstraight
        \fmftop{u1,u4}
        \fmfbottom{u2,u3}
        \fmf{phantom}{u1,u5,u3}
        \fmfdot{u1,u2,u3,u4,u5}
        \fmflabel{1}{u1}
        \fmflabel{2}{u2}
        \fmflabel{3}{u3}
        \fmflabel{4}{u4}
        \fmflabel{5}{u5}
        \fmffreeze
        \fmf{dashes,right=.4}{u1,u2,u3,u4,u1}
        \fmf{plain}{u1,u2}
        \fmf{plain}{u3,u5,u4}
\end{fmfgraph*}}}
\qquad\quad
\raisebox{0mm}[8mm][10mm]{\parbox{13mm}{
\begin{fmfgraph*}(10,10) \fmfstraight
        \fmftop{u1,u4}
        \fmfbottom{u2,u3}
        \fmf{phantom}{u1,u5,u3}
        \fmfdot{u1,u2,u3,u4,u5}
        \fmflabel{1}{u1}
        \fmflabel{2}{u2}
        \fmflabel{3}{u3}
        \fmflabel{4}{u4}
        \fmflabel{5}{u5}
        \fmffreeze
        \fmf{dashes,right=.4}{u1,u2,u3,u4,u1}
        \fmf{plain}{u1,u4}
        \fmf{plain}{u2,u5,u3}
\end{fmfgraph*}}}
\qquad\quad
\raisebox{0mm}[8mm][10mm]{\parbox{13mm}{
\begin{fmfgraph*}(10,10) \fmfstraight
        \fmftop{u1,u4}
        \fmfbottom{u2,u3}
        \fmf{phantom}{u1,u5,u3}
        \fmfdot{u1,u2,u3,u4,u5}
        \fmflabel{1}{u1}
        \fmflabel{2}{u2}
        \fmflabel{3}{u3}
        \fmflabel{4}{u4}
        \fmflabel{5}{u5}
        \fmffreeze
        \fmf{dashes,right=.4}{u1,u2,u3,u4,u1}
        \fmf{plain}{u1,u5,u4}
        \fmf{plain}{u2,u3}
\end{fmfgraph*}}}
\\
\qquad\quad
\raisebox{0mm}[10mm][8mm]{\parbox{13mm}{
\begin{fmfgraph*}(10,10) \fmfstraight
        \fmftop{u1,u4}
        \fmfbottom{u2,u3}
        \fmf{phantom}{u1,u5,u3}
        \fmfdot{u1,u2,u3,u4,u5}
        \fmflabel{1}{u1}
        \fmflabel{2}{u2}
        \fmflabel{3}{u3}
        \fmflabel{4}{u4}
        \fmflabel{5}{u5}
        \fmffreeze
        \fmf{dashes,right=.4}{u1,u2,u3,u4,u1}
        \fmf{plain}{u1,u2}
        \fmf{plain}{u3,u4}
\end{fmfgraph*}}}
\qquad\quad
\raisebox{0mm}[10mm][8mm]{\parbox{13mm}{
\begin{fmfgraph*}(10,10) \fmfstraight
        \fmftop{u1,u4}
        \fmfbottom{u2,u3}
        \fmf{phantom}{u1,u5,u3}
        \fmfdot{u1,u2,u3,u4,u5}
        \fmflabel{1}{u1}
        \fmflabel{2}{u2}
        \fmflabel{3}{u3}
        \fmflabel{4}{u4}
        \fmflabel{5}{u5}
        \fmffreeze
        \fmf{dashes,right=.4}{u1,u2,u3,u4,u1}
        \fmf{plain}{u1,u4}
        \fmf{plain}{u2,u3}
\end{fmfgraph*}}}
\end{gathered}
$$
\parbox{11cm}{
\caption{The four different planar contractions on the disc passing once through 
the fifth point (top line), and the two not passing through it (bottom line).
\label{fig-5-pt}}}
\end{center}
\end{figure}

Starting with our general insertion formula \eqn{4-pt}, and focusing 
on the terms where the interaction is among the points 
$x_1$, $x_2$, $x_3$ and $x_4$, one can examine all the remaining 
tree-level planar contractions on the disc. As in the case of the 
general four-point function it is clear that $D_{1234}$ will always 
show up in combinations of the form (see Figure~\ref{fig-5-pt})
\be
D_{1234}\big([12][34]+[14][23]\big)\,,\qquad
D_{1234}\big([12,34|5]+[14,23|5]\big)\,,\qquad
D_{1234}[15][25][35][45]\,.
\label{5-pt-const}
\ee
These terms will multiply some extra tree-level contractions and then 
one has to sum over the three inequivalent permutations of $1234$ and 
over the choice of other quadruples.

The last term in \eqn{5-pt-const} has no orientation, so we expect it to 
cancel once $D_{1243}$ and $D_{1324}$ are included. 
The two other terms in \eqn{5-pt-const} are the ingredients that make 
up the minimal four-point function 
\eqn{4pt-2222} and five-point function \eqn{5pt-22222} 
(note the relations in \eqn{4pt-2222-temp} and \eqn{V11112}). 
As mentioned, in the case of the general four-point function it turns 
out that the combinatorics are such that these terms exactly factorize 
as in \eqn{eins}.

For the general five-point function we do not have such a general 
proof, but for all the cases listed above, and for the those we computed 
in Appendix~\ref{app}, all the graphs can be reorganized in a 
simple manner. They are all given as the sum of the minimal 
five-point function \eqn{5pt-22222} and the five possible minimal 
four-point functions \eqn{4pt-2222}, each multiplied by certain 
tree-level contractions.

It would be interesting to check if this decomposition indeed holds 
beyond the examples computed in Appendix~\ref{app} and whether 
a similar structure persists to higher-loop order.

\section{Six-point functions}

Lastly we evaluate one six-point function
\be
\label{6pt-222222}
\vev{\cO^{u_1}_2\cO^{u_2}_2\cO^{u_3}_2\cO^{u_4}_2\cO^{u_5}_2\cO^{u_6}_2}_\text{1-loop}
\ee
All the disc amplitudes appearing in \eqn{4-pt} are equivalent, up to permutations 
of the indices. For example the amplitude multiplying $D_{1234}$ is
\be
\begin{aligned}
\vev{\cO^{u_1}_1\cO^{u_2}_1\cO^{u_3}_1\cO^{u_4}_1|\cO^{u_5}_2\cO^{u_6}_2}_\text{tree, disc}
=&\, 4\,\Big(([15][36]+[16][35]) ([25][46]+[26][45]) 
\\&\hskip-3cm
+ [12][56] ([36][45]+[35][46])+[14][56] ([25][36]+[26][35]) 
\\&\hskip-3cm
+[15][56] ([26][34]+[23][46]) + [16][56] ([25][34]+[23][45])\Big).
\end{aligned}
\label{V111122}
\ee
Note that now there are also two possible disconnected graphs 
$[56]^2([12][34]+[14][23])$, which we have not included.

As usual it is simpler to subtract all graphs, including non-planar ones, 
and get the result as minus the non-planar graphs
\be
\begin{aligned}
&-64D_{1234}\Big([13][56]\big([25][46]+[26][45]\big)
+[24][56]\big([15][36]+[16][35]\big)
\\&\hskip7cm
[15][35][26][46]+[16][36][25][45]\Big).
\end{aligned}
\ee
Summing over different permutations of these Wick-contractions with the 
relevant $D_{ijlm}$ factors gives the one-loop six-point amplitude. 

For simple six-point function we expect a similar decomposition as we have 
found in the sample five-point functions we have studied. Now the four boundary 
insertions on the disc can be contracted directly, giving terms like the 
minimal four-point function \eqn{4pt-2222}, through one of the internal 
vertices, giving terms proportional to the minimal five-point function 
\eqn{5pt-22222} or through both internal points giving the ingredients 
of the minimal six-point function \eqn{6pt-222222}. We do not know whether 
such considerations hold for arbitrary six-point functions, or for that matter 
for larger $n$-point functions.

\section{Extremal and next-to-extremal $n$-point functions}

Finally we consider a special class of $n$-point functions 
\be
\vev{\cO^{u_1}_{k_1} \cO^{u_2}_{k_2}\ldots  \cO^{u_{n-1}}_{k_{n-1}} \cO^{u_n}_{k} }
\qquad \text{with}\quad k=\sum_{i=1}^{n-1}k_i - m\,.
\label{extremal}
\ee
For $m=0$ these are known as the extremal and for $m=2$ as the next-to-extremal $n$-point functions,
which do not receive quantum corrections at one-loop order.%
\footnote{The case $m=1$ vanishes trivially.}
It is instructive to rederive this fact from \eqn{4-pt}. We have
\begin{align}
\vev{\cO^{u_1}_{k_1} \cO^{u_2}_{k_2}\ldots  \cO^{u_{n-1}}_{k_{n-1}} \cO^{u_n}_{k} }_\text{1-loop}=&\, \nn
\\&\hskip-3.5cm
\sum_{ijl} D_{ijln}\, k_i k_j k_l k
\vev{\cO^{u_i}_{k_i-1}\cO^{u_j}_{k_j-1}\cO^{u_l}_{k_l-1}\cO^{u_n}_{k-1}\Bigr|\prod_{p\neq i,j,l}\cO^{u_p}_{k_p}}_\text{tree, disc}
\label{peter}
\\&\hskip-3.5cm
+
\sum_{ijlm} D_{ijlm}\, k_i k_j k_l k_m
\vev{\cO^{u_i}_{k_i-1}\cO^{u_j}_{k_j-1}\cO^{u_m}_{k_l-1}\cO^{u_m}_{k_m-1}\Bigr |\cO^{u_n}_{k}
\prod_{p\neq i,j,l,m}\cO^{u_p}_{k_p}}_\text{tree, disc} \,, \nn
\end{align}
where the sums run over all orders of $ijl$ and $ijlm$ respectively.
Due to the special structure of \eqn{extremal} with $\cO^{u_n}_k$ having the largest dimension,
all the other $(n-1)$ operators $\cO^{u_i}_{k_i}$ of lower dimension have to contract with 
$\cO^{u_n}_k$. Hence the tree-level disc amplitudes
in the second line of \eqn{peter} vanish, as these contractions leave four 
($m=0$) or two ($m=2$) legs
of $\cO^{u_n}_k$ uncontracted. The same argument also kills the first sum in the extremal case ($m=0$),
which leaves two legs of  $\cO^{u_n}_k$ uncontracted.

In order to see the vanishing of the first sum in \eqn{peter} in the next-to-extremal case ($m=2$)
we note that the relevant tree-level disc amplitude
\be
\vev{\cO^{u_i}_{k_i-1}\cO^{u_j}_{k_j-1}\cO^{u_l}_{k_l-1}\cO^{u_n}_{k-1}\Bigr|\prod_{p\neq i,j,l}\cO^{u_p}_{k_p}}_\text{tree, disc}=[in]^{k_i-1}[jn]^{k_j-1}[kn]^{k_l-1}
\prod_{p\neq i,j,l}k_p [np]^{k_p}
\ee
is independent of the ordering of the labels $ijl$. Hence
\begin{align}
\label{paul}
\vev{\cO^{u_1}_{k_1} \cO^{u_2}_{k_2}\ldots  \cO^{u_{n-1}}_{k_{n-1}} \cO^{u_n}_{\sum_i k_i-2} }_\text{1-loop}=&\,
\\&\hskip-5cm
\sum_{(ijl)}  (D_{nijl}+ D_{njil}+D_{nilj})\, k_i k_j k_l k\, 
[in]^{k_i-1}[jn]^{k_j-1}[kn]^{k_l-1}
\prod_{p\neq i,j,l}k_p[np]^{k_p} =0\,,
\nn
\end{align}
where the sum is now over the ordered triples $(ijl)$ and the vanishing is due to the relation
\eqn{modular} for the $D$'s.

We therefore verified by our formalism the vanishing of the one-loop corrections 
to the extremal and next-to-extremal correlators.

\end{fmffile}

\subsection*{Acknowledgments}
We would like to thank Gleb Arutyunov, Niklas Beisert, Johannes Henn and
Donovan Young for stimulating discussions. We acknowledge 
the hospitality of the Galileo Galilei Institute in Firenze where part 
of this work was done and the INFN for partial financial support.
This work was supported by the Volkswagen Foundation.

\appendix

\section{A selection of one-loop five-point functions}
\label{app}

We collect here some more explicit results for the one-loop 
planar five-point functions of chiral primary operators. We only 
write down the one-loop part. The tree level is found by simple 
Wick-contractions. They are all expressed in terms of the one-loop 
four point function and five-point function of operators of dimension 
two \eqn{4pt-2222}, \eqn{5pt-22222}, 
multiplied by some tree level contractions. We do not write 
down explicitly the space-time point of each operator, the position 
$x_i$ is always matched with the index of $u_i$
\be
\begin{aligned}
\vev{\cO^{u_1}_2\cO^{u_2}_2\cO^{u_3}_2\cO^{u_4}_2\cO^{u_5}_6}_\text{1-loop}=0
\end{aligned}
\ee
\be
\begin{aligned}
\vev{\cO^{u_1}_2\cO^{u_2}_2\cO^{u_3}_2\cO^{u_4}_3\cO^{u_5}_5}
&=
15 [45]^3\vev{\cO^{u_1}_2\cO^{u_2}_2\cO^{u_3}_2\cO^{u_5}_2}
+\frac{45}{2} [35]^2[45]\vev{\cO^{u_1}_2\cO^{u_2}_2\cO^{u_4}_2\cO^{u_5}_2}
\\&\hskip-1cm
+\frac{45}{2} [25]^2[45]\vev{\cO^{u_1}_2\cO^{u_3}_2\cO^{u_4}_2\cO^{u_5}_2}
+\frac{45}{2} [15]^2[45]\vev{\cO^{u_2}_2\cO^{u_3}_2\cO^{u_4}_2\cO^{u_5}_2}
\end{aligned}
\ee
\begin{align}
\vev{\cO^{u_1}_2\cO^{u_2}_2\cO^{u_3}_2\cO^{u_4}_4\cO^{u_5}_4}
&=
4 [45]^2\vev{\cO^{u_1}_2\cO^{u_2}_2\cO^{u_3}_2\cO^{u_4}_2\cO^{u_5}_2}
+16 [34][35][45]\vev{\cO^{u_1}_2\cO^{u_2}_2\cO^{u_4}_2\cO^{u_5}_2}
\nn\\&\hskip-2cm
+16 [24][25][45]\vev{\cO^{u_1}_2\cO^{u_3}_2\cO^{u_4}_2\cO^{u_5}_2}
+16 [14][15][45]\vev{\cO^{u_2}_2\cO^{u_3}_2\cO^{u_4}_2\cO^{u_5}_2}
\end{align}
\be
\begin{aligned}
\vev{\cO^{u_1}_2\cO^{u_2}_2\cO^{u_3}_3\cO^{u_4}_3\cO^{u_5}_4}
&=
\frac{9}{2} [35][45]\vev{\cO^{u_1}_2\cO^{u_2}_2\cO^{u_3}_2\cO^{u_4}_2\cO^{u_5}_2}
\\&\hskip-1cm
+9 [34][45]^2\vev{\cO^{u_1}_2\cO^{u_2}_2\cO^{u_3}_2\cO^{u_5}_2}
+9 [34][35]^2\vev{\cO^{u_1}_2\cO^{u_2}_2\cO^{u_4}_2\cO^{u_5}_2}
\\&\hskip-1cm
+\frac{9}{2} [25] ( 2 [34][25] + [35][24] + [23][45] )
\vev{\cO^{u_1}_2\cO^{u_3}_2\cO^{u_4}_2\cO^{u_5}_2}
\\&\hskip-1cm
+\frac{9}{2} [15] ( 2 [34][15] + [35][14] + [13][45] )
\vev{\cO^{u_2}_2\cO^{u_3}_2\cO^{u_4}_2\cO^{u_5}_2}
\end{aligned}
\ee
\be
\begin{aligned}
\vev{\cO^{u_1}_2\cO^{u_2}_3\cO^{u_3}_3\cO^{u_4}_3\cO^{u_5}_3}
&=
\frac{81}{16} ([23][45] + [24][35] + [25][34])
\vev{\cO^{u_1}_2\cO^{u_2}_2\cO^{u_3}_2\cO^{u_4}_2\cO^{u_5}_2}
\\&\hskip-1cm
+\frac{81}{8} ([25,34|1] + [23,45|1] + [24,35|1])
\vev{\cO^{u_2}_2\cO^{u_3}_2\cO^{u_4}_2\cO^{u_5}_2}\,.
\end{aligned}
\ee
\be
\vev{\cO^{u_1}_2\cO^{u_2}_2\cO^{u_3}_2\cO^{u_4}_2\cO^{u_5}_7}=0
\ee
\be
\begin{aligned}
\vev{\cO^{u_1}_2\cO^{u_2}_2\cO^{u_3}_2\cO^{u_4}_4\cO^{u_5}_6}
&=
24[45]^4\vev{\cO^{u_1}_2\cO^{u_2}_2\cO^{u_3}_2\cO^{u_5}_2}
+48[34]^2[45]^2\vev{\cO^{u_1}_2\cO^{u_2}_2\cO^{u_4}_2\cO^{u_5}_2}
\\&\hskip-1cm
+48[24]^2[45]^2\vev{\cO^{u_1}_2\cO^{u_3}_2\cO^{u_4}_2\cO^{u_5}_2}
+48[14]^2[45]^2\vev{\cO^{u_2}_2\cO^{u_3}_2\cO^{u_4}_2\cO^{u_5}_2}
\end{aligned}
\ee
\begin{align}
\vev{\cO^{u_1}_2\cO^{u_2}_2\cO^{u_3}_3\cO^{u_4}_3\cO^{u_5}_6}
&=
\frac{81}{2}[34][45]^3\vev{\cO^{u_1}_2\cO^{u_2}_2\cO^{u_3}_2\cO^{u_5}_2}
+\frac{81}{2}[35]^3[45]\vev{\cO^{u_1}_2\cO^{u_2}_2\cO^{u_4}_2\cO^{u_5}_2}
\nn\\&\hskip-2.5cm
+54[25]^2[35][45]\vev{\cO^{u_1}_2\cO^{u_3}_2\cO^{u_4}_2\cO^{u_5}_2}
+54[15]^2[35][45]\vev{\cO^{u_2}_2\cO^{u_3}_2\cO^{u_4}_2\cO^{u_5}_2}
\end{align}
\begin{align}
\vev{\cO^{u_1}_2\cO^{u_2}_2\cO^{u_3}_2\cO^{u_4}_5\cO^{u_5}_5}
&=
\frac{25}{4}[45]^3\vev{\cO^{u_1}_2\cO^{u_2}_2\cO^{u_3}_2\cO^{u_4}_2\cO^{u_5}_2}
\nn\\&\hskip-1cm
+\frac{75}{2}[35][35][45]^2\vev{\cO^{u_1}_2\cO^{u_2}_2\cO^{u_4}_2\cO^{u_5}_2}
+\frac{75}{2}[25][25][45]^2\vev{\cO^{u_1}_2\cO^{u_3}_2\cO^{u_4}_2\cO^{u_5}_2}
\nn\\&\hskip-1cm
+\frac{75}{2}[15][15][45]^2\vev{\cO^{u_2}_2\cO^{u_3}_2\cO^{u_4}_2\cO^{u_5}_2}
\end{align}
\be
\begin{aligned}
\vev{\cO^{u_1}_2\cO^{u_2}_2\cO^{u_3}_3\cO^{u_4}_4\cO^{u_5}_5}
&=
\frac{15}{2}[35][45]^2\vev{\cO^{u_1}_2\cO^{u_2}_2\cO^{u_3}_2\cO^{u_4}_2\cO^{u_5}_2}
\\&\hskip-1cm
+30[34][45]^3 \vev{\cO^{u_1}_2\cO^{u_2}_2\cO^{u_3}_2\cO^{u_5}_2}
+30[35]^2[45][34]\vev{\cO^{u_1}_2\cO^{u_2}_2\cO^{u_4}_2\cO^{u_5}_2}
\\&\hskip-1cm
+\frac{15}{2} [25][45] (6 [25][34] + 4 [24][35] + 2 [23][45])
\vev{\cO^{u_1}_2\cO^{u_3}_2\cO^{u_4}_2\cO^{u_5}_2}
\\&\hskip-1cm
+\frac{15}{2} [15][45] (6 [15][34] + 4 [14][35] + 2 [13][45])
\vev{\cO^{u_2}_2\cO^{u_3}_2\cO^{u_4}_2\cO^{u_5}_2}
\end{aligned}
\ee
\begin{align}
\vev{\cO^{u_1}_2\cO^{u_2}_3\cO^{u_3}_3\cO^{u_4}_3\cO^{u_5}_5}
&=
\frac{135}{16}\vev{\cO^{u_1}_2\cO^{u_2}_2\cO^{u_3}_2\cO^{u_4}_2\cO^{u_5}_2}
\nn\\&\hskip-1cm
+\frac{135}{8}[45]^2 ( 2 [23][45] + [24][35]+ [25][34])
\vev{\cO^{u_1}_2\cO^{u_2}_2\cO^{u_3}_2\cO^{u_5}_2}
\nn\\&\hskip-1cm
+\frac{135}{8}[35]^2 ( 2 [24][35] + [23][45]+ [25][34])
\vev{\cO^{u_1}_2\cO^{u_2}_2\cO^{u_4}_2\cO^{u_5}_2}
\\&\hskip-1cm
+\frac{135}{8}[25]^2 ( 2 [25][34] + [24][35]+ [23][45])
\vev{\cO^{u_1}_2\cO^{u_3}_2\cO^{u_4}_2\cO^{u_5}_2}
\nn\\&\hskip-3cm
+\frac{135}{8}[15] (3 [15][25][34] + [12][35][45] + [14][25][35] + [13][25][45] )
\vev{\cO^{u_2}_2\cO^{u_3}_2\cO^{u_4}_2\cO^{u_5}_2}
\nn
\end{align}

\raggedright


\begin{thebibliography}{20}
\addtolength{\parskip}{-1ex}


\bibitem{AdSCFT}
J.~M.~Maldacena,
``The large $N$ limit of superconformal field theories and supergravity,''
Adv.\ Theor.\ Math.\ Phys.\  {\bf 2}, 231 (1998)
[Int.\ J.\ Theor.\ Phys.\  {\bf 38}, 1113 (1999)]
[hep-th/9711200]; \\
S.~S.~Gubser, I.~R.~Klebanov and A.~M.~Polyakov,
  ``Gauge theory correlators from non-critical string theory,''
  Phys.\ Lett.\  B {\bf 428} (1998) 105
  [hep-th/9802109]; \\
E.~Witten,
  ``Anti-de Sitter space and holography,''
  Adv.\ Theor.\ Math.\ Phys.\  {\bf 2}, 253 (1998)
  [hep-th/9802150].

\bibitem{AdS/INT}
  J.~A.~Minahan and K.~Zarembo,
  ``The Bethe-ansatz for ${\cal N}=4$ super Yang-Mills,''
  JHEP {\bf 0303}, 013 (2003)
  [hep-th/0212208];\\
  N.~Beisert, C.~Kristjansen and M.~Staudacher,
  ``The dilatation operator of ${\cal N}=4$ super Yang-Mills theory,''
  Nucl.\ Phys.\  B {\bf 664}, 131 (2003)
  [hep-th/0303060];\\
  I.~Bena, J.~Polchinski and R.~Roiban,
  ``Hidden symmetries of the $AdS_5\times  S^5$ superstring,''
  Phys.\ Rev.\  D {\bf 69}, 046002 (2004)
  [hep-th/0305116].

\bibitem{3pt}
  B.~Eden, P.~S.~Howe and P.~C.~West,
  ``Nilpotent invariants in ${\cal N}=4$ SYM,''
  Phys.\ Lett.\  B {\bf 463} (1999) 19
  [hep-th/9905085]; \\
G.~Arutyunov, B.~Eden and E.~Sokatchev,
  ``On non-renormalization and OPE in superconformal field theories,''
  Nucl.\ Phys.\  B {\bf 619}, 359 (2001)
  [hep-th/0105254]; \\
 P.~J.~Heslop and P.~S.~Howe,
  ``OPEs and 3-point correlators of protected operators in ${\cal N}=4$ SYM,''
  Nucl.\ Phys.\  B {\bf 626}, 265 (2002)
  [hep-th/0107212].

\bibitem{Lee:1998bxa}
S.~Lee, S.~Minwalla, M.~Rangamani and N.~Seiberg,
``Three-point functions of chiral operators in $D = 4$, $\cN = 4$ SYM at  large
$N$,''
Adv.\ Theor.\ Math.\ Phys.\  {\bf 2}, 697 (1998)
[hep-th/9806074].

\bibitem{Betheeq}
  N.~Beisert, R.~Hernandez and E.~Lopez,
  ``A crossing-symmetric phase for $AdS_5\times  S^5$ strings,''
  JHEP {\bf 0611}, 070 (2006)
  [hep-th/0609044];\\
  N.~Beisert, B.~Eden and M.~Staudacher,
  ``Transcendentality and crossing,''
  J.\ Stat.\ Mech.\  {\bf 0701}, P021 (2007)
  [hep-th/0610251].


\bibitem{intrevs}
A.~A.~Tseytlin,
   ``Semiclassical strings in $AdS_5\times  S^5$ and scalar operators in ${\cal N}=4$  SYM
  theory,''
  Comptes Rendus Physique {\bf 5}, 1049 (2004)
  [hep-th/0407218];\\
  N.~Beisert,
   ``The dilatation operator of ${\cal N}=4$ super Yang-Mills theory and
  integrability,''
  Phys.\ Rept.\  {\bf 405}, 1 (2005)
  [hep-th/0407277];\\
  K.~Zarembo,
  ``Semiclassical Bethe ansatz and $AdS$/CFT,''
  Comptes Rendus Physique {\bf 5}, 1081 (2004)
  [Fortsch.\ Phys.\  {\bf 53}, 647 (2005)]
  [hep-th/0411191];\\
  J.~Plefka,
   ``Spinning strings and integrable spin chains in the $AdS$/CFT
  correspondence,''
  Living Rev.\ Rel.\  {\bf 8}, 9 (2005)
  [hep-th/0507136];\\
  J.~A.~Minahan,
  ``A brief introduction to the Bethe ansatz In ${\cal N}=4$ super-Yang-Mills,''
  J.\ Phys.\ A  {\bf 39}, 12657 (2006).

\bibitem{GluonScattering}
 Z.~Bern, L.~J.~Dixon and V.~A.~Smirnov,
  ``Iteration of planar amplitudes in maximally supersymmetric Yang-Mills
  theory at three loops and beyond,''
  Phys.\ Rev.\  D {\bf 72} (2005) 085001
  [hep-th/0505205]; \\
 L.~F.~Alday and J.~M.~Maldacena,
  ``Gluon scattering amplitudes at strong coupling,''
  JHEP {\bf 0706} (2007) 064
  [arXiv:0705.0303];\\
J.~M.~Drummond, J.~Henn, G.~P.~Korchemsky and E.~Sokatchev,
  ``On planar gluon amplitudes/Wilson loops duality,''
  Nucl.\ Phys.\  B {\bf 795} (2008) 52
  [arXiv:0709.2368];\\
J.~M.~Drummond, J.~Henn, G.~P.~Korchemsky and E.~Sokatchev,
  ``Dual superconformal symmetry of scattering amplitudes in ${\cal N}=4$
  super-Yang-Mills theory,''
  arXiv:0807.1095;\\
N.~Berkovits and J.~Maldacena,
  ``Fermionic T-duality, dual superconformal symmetry, and the amplitude/Wilson
  loop connection,''
  JHEP {\bf 0809} (2008) 062
  [arXiv:0807.3196];\\
 N.~Beisert, R.~Ricci, A.~A.~Tseytlin and M.~Wolf,
  ``Dual superconformal symmetry from $AdS_5\times S^5$ superstring integrability,''
  arXiv:0807.3228.

\bibitem{AldayRoiban}
L.~F.~Alday and R.~Roiban,
  ``Scattering amplitudes, Wilson loops and the string/gauge theory
  correspondence,''
  Phys.\ Rept.\  {\bf 468} (2008) 153
  [arXiv:0807.1889];\\
J.~Henn, ``Dualit\'e entre boucles de Wilson et amplitudes de gluons'', PhD thesis,
Universit\'e Claude Bernard - Lyon 1.

\bibitem{Eden:1998hh}
B.~Eden, P.~S.~Howe, C.~Schubert, E.~Sokatchev and P.~C.~West,
``Four-point functions in $\cN = 4$ supersymmetric Yang-Mills theory at two
loops,''
Nucl.\ Phys.\  B {\bf 557} (1999) 355
[hep-th/9811172].

\bibitem{D'Hoker:1999pj}
  E.~D'Hoker, D.~Z.~Freedman, S.~D.~Mathur, A.~Matusis and L.~Rastelli,
  ``Graviton exchange and complete 4-point functions in the $AdS$/CFT
  correspondence,''
  Nucl.\ Phys.\  B {\bf 562}, 353 (1999)
  [hep-th/9903196].

\bibitem{D'Hoker:1999ni}
  E.~D'Hoker, D.~Z.~Freedman and L.~Rastelli,
  ``$AdS$/CFT 4-point functions: How to succeed at z-integrals without  really
  trying,''
  Nucl.\ Phys.\  B {\bf 562}, 395 (1999)
  [hep-th/9905049].

\bibitem{Eden:1999kh}
B.~Eden, P.~S.~Howe, C.~Schubert, E.~Sokatchev and P.~C.~West,
  ``Simplifications of four-point functions in ${\cal N}=4$ supersymmetric  Yang-Mills
  theory at two loops,''
  Phys.\ Lett.\  B {\bf 466} (1999) 20
  [hep-th/9906051].

\bibitem{Arutyunov:2000py}
G.~Arutyunov and S.~Frolov,
``Four-point functions of lowest weight CPOs in ${\cal N} = 4$ SYM$_4$ 
in  supergravity approximation,''
Phys.\ Rev.\  D {\bf 62} (2000) 064016
[hep-th/0002170].

\bibitem{Dolan:2000ut}
  F.~A.~Dolan and H.~Osborn,
  ``Conformal four point functions and the operator product expansion,''
  Nucl.\ Phys.\  B {\bf 599} (2001) 459
  [hep-th/0011040].

\bibitem{Heslop:2002hp}
  P.~J.~Heslop and P.~S.~Howe,
  ``Four-point functions in ${\cal N}=4$ SYM,''
  JHEP {\bf 0301} (2003) 043
  [hep-th/0211252].

\bibitem{Arutyunov:2002fh}
G.~Arutyunov, F.~A.~Dolan, H.~Osborn and E.~Sokatchev,
``Correlation functions and massive Kaluza-Klein modes in the $AdS$/CFT
correspondence,''
Nucl.\ Phys.\  B {\bf 665} (2003) 273
[hep-th/0212116].

\bibitem{Arutyunov:2003ae}
G.~Arutyunov and E.~Sokatchev,
``On a large $N$ degeneracy in $\cN = 4$ SYM and the $AdS$/CFT
correspondence,''
Nucl.\ Phys.\  B {\bf 663} (2003) 163
[hep-th/0301058].

\bibitem{Arutyunov:2003ad}
G.~Arutyunov, S.~Penati, A.~Santambrogio and E.~Sokatchev,
``Four-point correlators of BPS operators in $\cN=4$ SYM at order $g^4$,''
Nucl.\ Phys.\  B {\bf 670} (2003) 103
[hep-th/0305060].

\bibitem{Dolan:2004mu}
F.~A.~Dolan, L.~Gallot and E.~Sokatchev,
``On four-point functions of 1/2-BPS operators in general dimensions,''
JHEP {\bf 0409}, 056 (2004)
[hep-th/0405180].

\bibitem{Nirschl:2004pa}
M.~Nirschl and H.~Osborn,
``Superconformal Ward identities and their solution,''
Nucl.\ Phys.\  B {\bf 711} (2005) 409
[hep-th/0407060].

\bibitem{extremal}
 E.~D'Hoker, D.~Z.~Freedman, S.~D.~Mathur, A.~Matusis and L.~Rastelli,
``Extremal correlators in the $AdS$/CFT correspondence,''
  hep-th/9908160; \\
  B.~Eden, P.~S.~Howe, C.~Schubert, E.~Sokatchev and P.~C.~West,
  ``Extremal correlators in four-dimensional SCFT,''
  Phys.\ Lett.\  B {\bf 472} (2000) 323
  [hep-th/9910150]; \\
J.~Erdmenger and M.~Perez-Victoria,
  ``Non-renormalization of next-to-extremal correlators in ${\cal N}=4$ SYM and  the
  $AdS$/CFT correspondence,''
  Phys.\ Rev.\  D {\bf 62} (2000) 045008
  [hep-th/9912250]; \\
B.~U.~Eden, P.~S.~Howe, E.~Sokatchev and P.~C.~West,
  ``Extremal and next-to-extremal $n$-point correlators in four-dimensional
  SCFT,''
  Phys.\ Lett.\  B {\bf 494} (2000) 141
  [hep-th/0004102].

\bibitem{Arutyunov:2000im}
G.~Arutyunov, S.~Frolov and A.~Petkou,
``Perturbative and instanton corrections to the OPE of CPOs in $\cN = 4$
SYM$_4$,''
Nucl.\ Phys.\  B {\bf 602} (2001) 238
[Erratum-ibid.\  B {\bf 609} (2001) 540]
[hep-th/0010137].

\bibitem{us2}
N.~Drukker and J.~Plefka,
``Superprotected $n$-point correlation functions of local operators in $\cN=4$
super Yang-Mills,''
arXiv:0901.3653.


\bibitem{Beisert:2002bb}
  N.~Beisert, C.~Kristjansen, J.~Plefka, G.~W.~Semenoff and M.~Staudacher,
  ``BMN correlators and operator mixing in $\cN = 4$ super Yang-Mills theory,''
  Nucl.\ Phys.\  B {\bf 650} (2003) 125
  [hep-th/0208178].


\bibitem{David}
  N.~I.~Usyukina and A.~I.~Davydychev,
  ``An Approach to the evaluation of three and four point ladder diagrams,''
  Phys.\ Lett.\  B {\bf 298} (1993) 363.

\bibitem{nearext}
E.~D'Hoker, J.~Erdmenger, D.~Z.~Freedman and M.~Perez-Victoria,
  ``Near-extremal correlators and vanishing supergravity couplings in
  $AdS$/CFT,''
  Nucl.\ Phys.\  B {\bf 589} (2000) 3
  [hep-th/0003218].








\end{thebibliography}
\end{document}